\documentclass[preprint]{revtex4}
\usepackage{graphicx}
\begin{document}
\preprint{}
\title{On growth of spinodal instabilities in nuclear matter-II:asymmetric matter}
\author{F. Acar$^{1}$}
\author{S. Ayik$^{2,}$}\email{ayik@tntech.edu}
\author{O. Yilmaz$^{1,}$}\email{oyilmaz@metu.edu.tr}
\author{A. Gokalp$^{3}$}

\affiliation{$^{1}$Physics Department, Middle East Technical
University, 06800 Ankara, Turkey \\
$^{2}$Physics Department, Tennessee Technological University,
Cookeville, TN 38505, USA \\
$^{3}$Department of Physics, Bilkent University, 06800 Ankara,
Turkey }
\date{\today}

\begin{abstract}
 As an extension of our previous work, the growth of density fluctuations in the spinodal region of charge asymmetric nuclear matter is investigated in the basis of the stochastic mean-field approach in the non-relativistic framework. A complete treatment of density correlation functions are presented by including collective modes and non-collective modes as well.
\end{abstract}
\maketitle

\section{Introduction}

 Nuclear matter in the spinodal region exhibits a universal behavior. When the uniform matter enters into the spinodal region, small fluctuations grow rapidly and the matter undergoes a dynamical liquid-gas phase transformation. Nuclear multi-fragmentation processes are considered as a signature for such a dynamical phase transformation in nuclear matter [1-3]. Theoretical description of the spinodal dynamics requires an approach beyond the mean-field theory by incorporating fluctuation mechanisms [4-7]. The recently proposed stochastic mean-field approach (SMF) provides a useful framework for describing large amplitude nuclear collective motion by including one-body dissipation and fluctuation mechanisms [8]. We carried out a number of investigations of the spinodal instabilities by employing the non-relativistic and relativistic framework of the SMF approach [9-13]. For a full description of the spinodal dynamics, we need to determine an ensemble of events by simulating the SMF equation with proper fluctuations in the initial state. In our investigation, instead of such simulations, we calculate the early growth of density fluctuations in the linear response regime of the SMF approach. For this purpose, the equal time correlation function of the density fluctuations provides a very useful quantity.  For nuclear matter, in the linear response regime, it is possible to provide a nearly analytical treatment for the density correlation function. Indeed, the density correlation function exhibits the early stages of the dynamics of the liquid-gas phase transformation. In our earlier studies, we calculated the correlation function in the pole approximation. As pointed out by Bozek, the effect of non-collective poles play an important role in the full description of the correlation function [14]. In a recent work, we calculated the density correlation function exactly by including the effects of non-collective poles in addition to the collective poles for charge symmetric nuclear matter [15]. In this work, we present calculations of the density correlation function for charge asymmetric nuclear matter in the semi-classical framework, i.e. in the long wavelength limit, of the SMF approach. In section 2, we briefly present description of the density correlation functions including collective and non-collective modes in the linear response framework in the long wavelength limit. In section 3, we present calculations for the most unstable initial density $\rho =0.3\rho _{0} $ at two different temperatures and two charge asymmetries of the matter. Conclusions are given in section 4.

\section{ Density correlation functions in asymmetric nuclear matter}

 A detailed description of the correlation function of the density fluctuations for symmetric matter including the non-collective poles is presented in our recent publication [15]. Description of the correlation function for the charge asymmetric system is given in a similar manner. Here, we give a brief description and refer to [9,15] for details. For charge asymmetric case, we need to consider the coupled SMF equations for neutrons and protons. The linearized equations for fluctuations of neutron and proton density matrices, $\delta \hat{\rho }_{a}^{\lambda } (t)=\hat{\rho }_{a}^{\lambda } (t)-\hat{\rho }_{a} $  are given by
\begin{equation} \label{Eq1}
i\hbar \frac{\partial }{\partial t} \delta \hat{\rho }_{a}^{\lambda } (t)=\left[h_{a} ,\delta \hat{\rho }_{a}^{\lambda } (t)\right]+\left[\delta U_{a}^{\lambda } (t),\hat{\rho }_{a} \right],
\end{equation}
where $\hat{\rho}_{a} $ is the density matrix of the initial state, $h_{a} $ denotes the mean-field Hamiltonian at the initial state and $\delta U_{a}^{\lambda} $ is the fluctuating part of the mean-field for protons and neutrons $a=p,n$ in the event labeled by $\lambda $. Since, for infinite matter the equilibrium state and the mean-field Hamiltonian are homogeneous, it is suitable to analyze these equations in plane wave representations. Then, the linear response equation becomes
\begin{equation} \label{Eq2}
\begin{array}{l} {i\hbar \frac{\partial }{\partial t} <\vec{p}_{1} |\delta \hat{\rho }_{a}^{\lambda } (t)|\vec{p}_{2} >} \\ {=\left[\varepsilon _{a} (\vec{p}_{1} )-\varepsilon_{a} (\vec{p}_{2} )\right]<\vec{p}_{1} |\delta \hat{\rho }_{a}^{\lambda } (t)|\vec{p}_{2} >-\left[\rho_{a} (\vec{p}_{1} )-\rho_{a} (\vec{p}_{2} )\right]<\vec{p}_{1} |\delta U_{a}^{\lambda} (t)|\vec{p}_{2} >} \end{array}
\end{equation}
The space Fourier transform of nucleon density fluctuations  $\delta \rho^{\lambda} (\vec{k},t)$ is related to the fluctuations of the density matrix according to
\begin{equation} \label{Eq3}
\delta {\rho }_{a} (\vec{k},t)=\sum _{s}\int \frac{d^{3} p}{(2\pi \hbar)^{3}}  <\vec{p}+\hbar \vec{k}/2|\delta \hat{\rho }_{a,s} (t)|\vec{p}-\hbar \vec{k}/2> ,
\end{equation}
where the summation indicates sum over spin quantum number $s=\uparrow ,\downarrow$. We solve the linear response Eq. (\ref{Eq2}) by employing the method of one-sided Fourier transformation in time [16-17], and obtain coupled  algebraic equations for the Fourier transform of the local proton and neutron densities
$\delta {\rho }_{a}^{\lambda } (\vec{k},\omega )$,
\begin{equation} \label{Eq4}
\begin{array}{l} {} \\ {\left(\begin{array}{c} {\left[1+F_{0}^{nn} \chi _{n} \left(\vec{k},\omega \right)\right]\delta {\rho }_{n}^{\lambda} (\vec{k},\omega )+F_{0}^{np} \chi _{n} \left(\vec{k},\omega \right)\delta {\rho }_{p}^{\lambda } (\vec{k},\omega )} \\ {\left[1+F_{0}^{pp} \chi _{p} \left(\vec{k},\omega \right)\right]\delta {\rho }_{p}^{\lambda } (\vec{k},\omega )+F_{0}^{pn} \chi _{p} \left(\vec{k},\omega \right)\delta {\rho }_{n}^{\lambda } (\vec{k},\omega )} \end{array}\right)=i\left(\begin{array}{c} {S_{n}^{\lambda } \left(\vec{k},\omega \right)} \\ {S_{p}^{\lambda } \left(\vec{k},\omega \right)} \end{array}\right)} \end{array}.
\end{equation}
In deriving these coupled equations, we consider that the mean-field potential depends only on the local nucleon densities, $U_{a}^{\lambda} =U\left(\rho _{p}^{\lambda} ,\rho _{n}^{\lambda} \right)$. The source term $S_{a}^{\lambda} \left(\vec{k},\omega \right)$ is determined by the matrix elements of fluctuations of the initial density matrix $\delta \hat{\rho}_{a,s}^{\lambda} (0)$ in spin-isospin channel as,
\begin{equation} \label{Eq5}
S_{a}^{\lambda}(\vec{k},\omega)=\sum_{s}\int \frac{d^{3} p}{(2\pi\hbar )^{3}} \frac{<\vec{p}+\hbar\vec{k}/2|\delta \hat{\rho}_{a,s}^{\lambda} (0)|\vec{p}-\hbar \vec{k}/2>}{\vec{v}\cdot \vec{k}-\omega}.
\end{equation}
According to the basic postulate of the SMF approach, elements of the initial density matrix are uncorrelated Gaussian random numbers with zero mean values and with well-defined variances. In semi-classical limit their variances are given by,
\begin{equation} \label{Eq6}
\begin{array}{l}{\overline{<\vec{p}+\hbar \vec{k}/2|\delta\hat{\rho}_{a,s}^{\lambda}(0)|\vec{p}-\hbar \vec{k}/2><{\vec{p}}^{\prime}-\hbar \vec{k}'/2|\delta \hat{\rho}_{b,s'}^{\lambda}(0)|\vec{p}'+\hbar\vec{k}'/2>}} \\ {=\delta _{ab} \delta_{ss'} (2\pi\hbar)^{6}\delta(\vec{p}-\vec{p}')\delta(\hbar\vec{k}-\hbar\vec{k}')f_{a} (\vec{p})\left(1-f_{a} (\vec{p})\right)} \\ {} \end{array}.
\end{equation}
The factor $\delta _{ab} \delta _{ss'} $ reflects the assumption that local density fluctuations in spin-isospin channels are uncorrelated in the initial state. In Eq. (\ref{Eq4}), $F_{0}^{ab}$ and $\chi_{a} \left(\vec{k},\omega \right)$ denote the zeroth order Landau parameters  and the Lindhard functions for protons and neutrons, respectively.  Landau parameters are defined by the derivative of the mean-field potential energy with respect to the proton and neutron densities evaluated at the initial state, $F_{0}^{ab} =\left(\partial U_{b} /\partial \rho _{a} \right)_{0}$. The semi-classical expression of the Lindhard functions for neutrons and protons are given by
\begin{equation} \label{Eq7}
\chi _{a} (\vec{k},\omega )=2\int \frac{d^{3} p}{(2\pi {\rm \; }\hbar )^{3} }  \frac{\vec{p}\cdot \hbar \vec{k}/m}{\hbar \omega -\vec{p}\cdot \hbar \vec{k}/m} \frac{\partial f_{a} (p)}{\partial \varepsilon } .
\end{equation}
Here the factor $2$ comes from spin summation, and $f_{a} (p)$ is the Fermi-Dirac factor for protons or neutrons at the initial state.

 According to the method of one-sided Fourier transform [16-17], we can calculate time evolution of the local density fluctuations by inverting Eq. (\ref{Eq4}) and taking the inverse Fourier transform to give,
\begin{equation} \label{Eq8}
\delta {\rho }_{a}^{\lambda} (\vec{k},t)=-i\int _{-\infty +i\sigma }^{+\infty +i\sigma }\frac{d\omega }{2\pi }  \frac{G_{a}^{\lambda } (\vec{k},\omega )}{\varepsilon (\vec{k},\omega )} e^{-i\omega t} ,
\end{equation}
where the integral in the complex $\omega $ is carried out along a contour that passes above all singularities of the integrand. The quantity $G_{a}^{\lambda} (\vec{k},\omega)$ in the integrand is given for neutrons and protons as
\begin{equation} \label{Eq9}
\left(\begin{array}{c} {G_{n}^{\lambda } \left(\vec{k},\omega \right)} \\ {G_{p}^{\lambda } \left(\vec{k},\omega \right)} \end{array}\right)=\left(\begin{array}{c} {\left[1+F_{0}^{pp} \chi _{p} \left(\vec{k},\omega \right)\right]S_{n}^{\lambda } \left(\vec{k},\omega \right)-F_{0}^{np} \chi _{n} \left(\vec{k},\omega \right)S_{p}^{\lambda } \left(\vec{k},\omega \right)} \\ {\left[1+F_{0}^{nn} \chi _{n} \left(\vec{k},\omega \right)\right]S_{p}^{\lambda } \left(\vec{k},\omega \right)-F_{0}^{pn} \chi _{p} \left(\vec{k},\omega \right)S_{n}^{\lambda } \left(\vec{k},\omega \right)} \end{array}\right) ,
\end{equation}
and $\varepsilon \left(\vec{k},\omega \right)$ denotes the susceptibility
\begin{equation} \label{Eq10}
\varepsilon \left(\vec{k},\omega \right)=1+F_{0}^{nn} \chi _{n} \left(\vec{k},\omega \right)+F_{0}^{pp} \chi _{p} \left(\vec{k},\omega \right)+\left[F_{0}^{nn} F_{0}^{pp} -F_{0}^{np} F_{0}^{pn} \right]\chi _{n} \left(\vec{k},\omega \right)\chi _{p} \left(\vec{k},\omega \right).
\end{equation}
As discussed in Ref. [15], it is possible to calculate the $\omega$-integral in Eq. (\ref{Eq8}), by choosing a contour as shown  in Fig. 1 of Ref. [15]. There are collective poles determined by the roots of the dispersion relation, $\varepsilon (\vec{k},\omega )=0\to \omega =\pm i\Gamma _{k}$.  The collective poles play important role in early growth of density fluctuations in the spinodal region. However, the collective poles alone do not provide a complete description of the growth of density fluctuations. The collective poles alone do not even satisfy the initial conditions as pointed in [14]. For a complete description, in addition to the collective poles, the effect of non-collective poles should be included into integral in Eq. (\ref{Eq8}). By calculating the angular integration in the Lindhard functions, it is easy to see that the integrand has a logarithmic singularity. The integrand in complex $\omega$-plane is multivalued, therefore entire real $\omega $-axis is a branch cut. In order to calculate integral in Eq. (\ref{Eq8}), we choose the contour  $C$, as shown in Fig. 1 of Ref. [15]. We exclude the real $\omega $-axis by drawing the contour from $+\infty $ to the origin just above the real $\omega $-axis, and after jumping from the first Riemann sheet to the second Riemann sheet at the origin, drawing contour just below the real $\omega $-axis from origin to $+\infty $. Contour is completed with a large semi-circle and by jumping from the second Riemann sheet to the first one at origin. As a result, we find that the integral in Eq. (\ref{Eq8}) can be expressed as
\begin{equation} \label{Eq11}
\delta {\rho }_{a}^{\lambda } (\vec{k},t)=\delta {\rho }_{a}^{\lambda } (P;\vec{k},t)+\delta {\rho }_{a}^{\lambda } (C;\vec{k},t),
\end{equation}
where the pole (P) contribution and cut (C) contribution are given by
\begin{equation} \label{Eq12}
\delta {\rho }_{a}^{\lambda} (P;\vec{k},t)=-\sum_{\pm}\frac{G_{a}^{\lambda} (\vec{k},\pm i\Gamma_{k})}{\partial \varepsilon (\vec{k},\omega )/\partial \omega |_{\omega =\pm i\Gamma_{k}}} e^{\pm \Gamma_{k} t} ,
\end{equation}
and
\begin{equation} \label{Eq13}
\delta {\rho }_{a}^{\lambda } (C;\vec{k},t)=-i\int _{-\infty }^{+\infty }\frac{d\omega }{2\pi } \left[\frac{G_{a}^{\lambda } (\vec{k},\omega +i\eta )}{\varepsilon (\vec{k},\omega +i\eta )} -\frac{G_{a}^{\lambda } (\vec{k},\omega -i\eta )}{\varepsilon (\vec{k},\omega -i\eta )} \right] e^{-i\omega t} .
\end{equation}

In order to investigate development of spinodal instabilities a very useful quantity is the equal time correlation functions of local density fluctuations
\begin{equation} \label{Eq14}
\sigma _{ab} (|\vec{r}-\vec{r}'|,t)=\overline{\delta \rho _{a}^{\lambda} (\vec{r},t)\delta \rho_{b}^{\lambda} (\vec{r}',t)}=\int \frac{d^{3} k}{(2\pi )^{3}}  e^{i\vec{k}\cdot (\vec{r}-\vec{r}')} \sigma_{ab} (\vec{k},t) .
\end{equation}
Here the local nucleon density fluctuations $\delta \rho_{a}^{\lambda} (\vec{r}',t)$are determined by the Fourier transform of  $\delta {\rho }_{a}^{\lambda } (\vec{k},t)$, and the spectral intensity $\sigma _{ab} (\vec{k},t)$of the correlation function is defined in terms of the variance of the Fourier transform of density fluctuations according to
\begin{equation} \label{Eq15}
\sigma _{ab} (\vec{k},t)(2\pi )^{3} \delta (\vec{k}-\vec{k}')=\overline{\delta {\rho }_{a}^{\lambda } (\vec{k},t)\delta {\rho }_{b}^{\lambda } (-\vec{k}',t)}  .
\end{equation}
In these expression, the bar indicates the average taken over the ensemble generated from the distribution of the initial fluctuations.

We can calculate the spectral intensity $\sigma_{ab} (\vec{k},t)$ by evaluating the ensemble averages using Eqs. (\ref{Eq12}) and (\ref{Eq13}) for pole $\delta {\rho}_{a}^{\lambda} (P;\vec{k},t)$ and cut $\delta {\rho}_{a}^{\lambda} (C;\vec{k},t)$ part of the Fourier transform density fluctuations and the Eq. (\ref{Eq6}) for the initial fluctuations. As a result, the spectral intensity is expressed as,
\begin{equation} \label{Eq16}
\sigma_{ab} (\vec{k},t)=\sigma_{ab} (PP;\vec{k},t)+2\sigma_{ab} (PC;\vec{k},t)+\sigma_{ab} (CC;\vec{k},t),
\end{equation}
where the first and last term are due to pole and cut parts of the spectral intensity and the middle term denotes the mixed contribution. We briefly describe derivation of the analytical expressions of the various terms in the spectral intensity in the long wavelength limit in Appendix A. The total spectral intensity is obtained by summing over isospin components
\begin{equation} \label{Eq17}
\sigma (\vec{k},t)=\sigma _{pp} (\vec{k},t)+2\sigma _{pn} (\vec{k},t)+\sigma _{nn} (\vec{k},t).
\end{equation}
The expression for the total correlation function of density fluctuations $\sigma (|\vec{r}-\vec{r}'|,t)$, which is summed over isospin components, is determined by using the total spectral density $\sigma (\vec{k},t)$ in Eq. (\ref{Eq14}). Using Eq. (\ref{Eq6}), we can determine the initial condition of the spectral density $\sigma (\vec{k},0)$, which must be equal to the right hand side of Eq. (\ref{Eq18}) at time $t=0$. This leads to a non-trivial sum rule
\begin{equation} \label{Eq18}
\sum _{p,n}\int 2\frac{d^{3} p}{(2\pi \hbar )^{3} }  f_{a} (\vec{p})\left(1-f_{a} (\vec{p})\right) =\sigma _{pp} (\vec{k},0)+2\sigma _{pn} (\vec{k},0)+\sigma _{nn} (\vec{k},0).
\end{equation}

\section{Results of calculations}

In numerical calculations, we employ the effective Skyrme potential of Ref. [4]
\begin{equation} \label{Eq19}
U_{a} (\rho_{n} ,\rho_{p} )=A\left(\frac{\rho}{\rho_{0}} \right)+B\left(\frac{\rho}{\rho_{0}} \right)^{\alpha+1} +C\left(\frac{\rho'}{\rho_{0}} \right)\tau_{a} +\frac{1}{2} \frac{dC}{d\rho} \frac{\rho'^{2} }{\rho_{0}} -D\Delta\rho +D'\Delta \rho'\tau_{a}
\end{equation}
where $\rho =\rho_{n} +\rho_{p} $ and $\rho'=\rho _{n}-\rho_{p}$ are the total and relative densities, $\tau _{a} =+1$ for neutrons and $\tau _{a} =-1$ for protons. The parameters $A=-356.8$ MeV, $B=+303.9$ MeV, $\alpha=1/6$ and $D=+130.0$ MeV $\text{fm}^{5}$ are adjusted  to reproduce the saturation properties of nuclear matter: the binding energy $\varepsilon_{0}=15.7$ MeV per nucleon, zero pressure at the saturation density $\rho_{0}=0.16$ ${\text{fm}}^{-3}$, compressibility coefficient $K=201$ MeV and the surface energy coefficient in the Weizsacker mass formula $a_{surf} =18.6$ MeV.  Magnitude of the parameter $D'=+34$ MeV ${\text{fm}}^{5}$ is close to the magnitude given by the $SkM^{*}$ [18]. The symmetry energy coefficient is $C(\rho )=C_{1} -C_{2} (\rho /\rho_{0})^{\alpha}$ with $C_{1}=+124.9$ MeV and $C_{2} =+93.5$ MeV.  These values for the symmetry energy coefficient in the Weizsacker mass formula gives $a_{sym} =\varepsilon_{F} (\rho_{0})/3+C(\rho_{0})/2=36.9/3+31.4/2=28.0$ MeV. We define the initial charge asymmetry as $I=\left(\rho_{n}-\rho_{n}\right)/\rho $.

Fig.1 shows phase diagrams for different charge asymmetries corresponding to different wavelengths in temperature-density plane. These diagrams indicate the boundaries of spinodal unstable regions for different wavelengths, starting from upper most boundary for $\lambda =\infty $.  The critical temperatures corresponding to this effective Skyrme potential depends on the initial charge asymmetries $I=0.0$, $I=0.4$ and $I=0.8$ are given
by $T_{c} =15$ MeV, $T_{c} =14$ MeV and $T_{c}=10$ MeV, respectively. The critical temperatures occur approximately around the same initial density $\rho=0.3\rho_{0}$. We notice that, consistent with earlier calculations [1], the spinodal instability region shrinks for increasing values of charge asymmetry. Neutron-rich matter with $I=0.8$ and $T=1.0$ MeV, approximately correspond to the structure of the crust of neutron stars. Under these conditions, the limiting spinodal boundary occurs at nucleon density around $\rho =0.55\rho_{0} $, which is consistent with the result found in [19]. We pick a reference state located at the center of the spinodal region with an initial density $\rho=0.3\rho_{0} $, and calculate the equal time correlation function of density fluctuations at two different temperatures $T=1.0$ MeV and $T=5.0$ MeV.  In Figs. 2-3, we plot the total spectral intensity $\sigma (\vec{k},t)$ of correlation functions as a function of wave number $k$ at time $t=40$ fm/$c$ for two different temperatures and three different initial charge asymmetries. At each initial charge asymmetry and temperatures, the upper limit of the wave number range $k_{\max }$ is determined by the condition that the inverse growth rate of the mode vanishes, $\Gamma _{k} =0$.  Dashed, dash-dotted and solid lines indicate the result of calculations in Eq. (\ref{Eq17}) with pole contributions $\sigma{}_{ab} (PP;\vec{k},t)$ only, with cut contributions only and the total of all terms, respectively. The cut contributions include cut-cut contribution $\sigma {}_{ab} (CC;\vec{k},t)$ and the cross terms due to pole-cut parts $2\sigma_{ab} (PC;\vec{k},t)$. From these figures, similar to the charge symmetric matter, we make two important observations for each value of the charge asymmetry: Cut terms make an important negative contribution during the early phase of growth, hence slowing down the growth of instabilities. During later times, collective poles dominate the growth of density fluctuations, and the cut terms representing the effects of non-collective poles do not grow in time, as discussed in earlier studies [1]. Second point that we note, both pole and cut contributions have divergent behavior with opposite sign, as wave numbers approach its upper limit, $k\to k{}_{\max}$. These divergent behaviors cancel out each other to produce a nice regular behavior of the spectral intensity as a function of wave number. Fig. 4 shows the total spectral intensity of the correlation function $\sigma (\vec{k},t)$ for a typical charge asymmetry $I=0.4$ as a function of wave number $k$ at different times for initial temperatures $T=1.0$ MeV and $T=5.0$ MeV. We observe that the spectral intensity, which includes pole and cut contributions, clearly exhibits the growth of the initial density fluctuations. The initial value of the spectral intensity at time $t=0$ indicated by the solid line is obtained by numerical calculations of the pole-pole terms, $\sigma{}_{ab} (PP;\vec{k},t=0)$, cut-cut terms $\sigma{}_{ab} (CC;\vec{k},t=0)$ and pole-cut terms $2\sigma_{ab} (PC;\vec{k},t=0)$.  On the other hand, the solid circles on this line are calculated by the initial conditions located on the left hand side of Eq. (\ref{Eq18}).  Nice agreement of both sides reflects the validity of highly non-trivial sum rule presented by Eq. (\ref{Eq18}).

 Fig. 5 shows the equal time correlation functions $\sigma(|\vec{r}-\vec{r}'|,t)$ of the total density fluctuations as a function distance of two space locations $x=|\vec{r}-\vec{r}'|$ at the initial density $\rho =0.3\rho_{0}$ and temperature $T=1.0$ MeV for three different charge asymmetries. The density correlation functions are calculated with complete spectral intensities shown by solid lines in Fig. 5, and the results are plotted at three different times $t=20,30,40$ fm/$c$. In Fig. 6, these correlation functions are plotted at temperature $T=5.0$ MeV. The growth and organization of the correlation function describes the initial phase of the condensation mechanism. In order to recognize the condensation mechanism, it is useful to introduce the correlation length $x_{C}$ as the width of the correlation function at half maximum. The correlation length provides a measure for size of condensing droplets during growth of fluctuations. In the correlation volume $\Delta V_{C} =4\pi x_{C}^{3}/3$, the variance of local density fluctuations at time $t$ is approximately given by $\sigma(x_{C},t)$. The number of nucleons in each correlation volume fluctuates with a dispersion $\Delta A_{C} =\Delta V_{C} \sqrt{\sigma (x_{C} ,t)}$. As a result, the total nucleon number in each correlation volume fluctuates approximately within the range $\Delta A_{0} -\Delta A_{C} \le \Delta A\le \Delta A_{0} +\Delta A_{C}$, where $\Delta A_{0} =\Delta V_{C}\rho_{0}$ denotes the number of nucleons at the initial uniform state. As seen from Figures 5 and 6, the correlation length is not very sensitive to the temperature and time evolution, but depends on the initial charge asymmetry. For example, at temperature $T=5.0$ MeV and the initial charge asymmetry $I=0.4$, the correlation length is about $x_{C}=3.0$ fm. In this case, the magnitude of the dispersion of density fluctuations at time $t=30$ fm/$c$ is about $\sqrt{\sigma (x_{C} ,t)} =0.04$ ${\text{fm}}^{-3}$. Consequently, the number of nucleons in the correlation volume approximately fluctuates in the range of $1\le \Delta A\le 9$. From this analysis, we observe that the spinodal decomposition indeed provides a dynamical mechanism for the liquid-gas phase transition. In the linear response approach, we can only recognize the early phase of the transition. In order to describe the full phase transition, we need to carry out long time simulations, which is not done in this study.

\section {Conclusions}

\noindent As a continuation of a previous work, we investigate early growth of density fluctuations in charge asymmetric nuclear matter in the basis of the SMF approach. In the linear response framework, employing the method of one-sided Fourier transform, it is possible to carry out nearly analytical treatment of the correlation function of density fluctuations. The density correlation function provides very useful information about the initial phase of the liquid-gas phase transformation of the system. The method of one-sided Fourier transform involves a contour integration in the complex frequency plane. In earlier investigations, this contour integral is evaluated by keeping only effects of the collective poles associated with unstable collective modes. This approximation, although it describes the growth of density fluctuations in a satisfactory manner, has important drawbacks.  It does not satisfy the initial conditions and furthermore leads to a divergent behavior as the wave numbers approach to their upper limits. In this work, we consider charge asymmetric nuclear matter in non-relativistic framework of the SMF approach.  We calculate the correlation function of density fluctuations including the effects of collective poles and non-collective poles in terms of the cut contribution in the complex frequency plane.  The cut contribution slows down the exponential growth of the pole contributions. It has also a divergent behavior with opposite sign, as wave numbers approach their upper limits. As a result divergent behaviors of pole and contributions cancel out each other to produce a nice regular behavior of the spectral intensity as a function of wave number. This allows us to have a complete description of the correlation function of density fluctuations in the linear response regime. Furthermore, exact calculations of the correlation functions satisfy the initial conditions expressed by a highly non-trivial sum rule.

\begin{acknowledgments}

S.A. gratefully acknowledges TUBITAK and the Middle East Technical University for partial support and warm hospitality extended to him during his visits. This work is supported in part by US DOE Grant No. DE-FG05-89ER40530, and in part by TUBITAK Grant No. 114F151.

\end{acknowledgments}

\appendix
\section{}

As seen in Eq. (\ref{Eq17}), there are three parts in the spectral intensity $\sigma _{ab} (\vec{k},t)$, the pole-pole part, the cut-cut part and the mixed cut-pole parts.The pole-pole part is
\begin{equation} \label{EqA1}
 \tilde{\sigma }_{ab} (PP;\vec{k},t)=\frac{E_{ab}^{+} }{|\left[\partial \varepsilon (\vec{k},\omega )/\partial \omega \right]_{\omega =i\Gamma _{k} } |^{2} } \left(e^{+2\Gamma _{k} t} +e^{-2\Gamma _{k} t} \right)+\frac{2E_{ab}^{-} }{|\left[\partial \varepsilon (\vec{k},\omega )/\partial \omega \right]_{\omega =i\Gamma _{k} } |^{2} } ,
   \end{equation}
where the quantities $E_{ab}^{\mp } $ for neutrons and protons are given by

\begin{equation} \label{EqA2}
 E_{nn}^{\mp } =\left[1+F_{0}^{pp} \chi _{p} \right]^{2} I_{n}^{\mp } +\left[F_{0}^{np} \chi _{n} \right]^{2} I_{p}^{\mp },
\end{equation}

\begin{equation}  \label{EqA3}
 E_{pp}^{\mp } =\left[1+F_{0}^{nn} \chi _{n} \right]^{2} I_{p}^{\mp } +\left[F_{0}^{pn} \chi _{p} \right]^{2} I_{n}^{\mp },
\end{equation}
and

\begin{equation}  \label{EqA4}
E_{np}^{\mp } =-\left(1+F_{0}^{pp} \chi _{p} \right)F_{0}^{pn} \chi _{p} I_{n}^{\mp } -\left(1+F_{0}^{nn} \chi _{n} \right)F_{0}^{np} \chi _{n} I_{p}^{\mp },
\end{equation}
with
\begin{equation}\label{EqA5}
I_{a}^{\mp} = 2\int \frac{d^{3} p}{(2\pi {\rm \; }\hbar )^{3} }  \frac{\left[\left(\Gamma _{k} \right)^{2} \mp (\vec{p}\cdot \vec{k}/m)^{2} \right]}{\left[(\Gamma _{k} )^{2} +\left(\vec{p}\cdot \vec{k}/m\right)^{2} \right]^{2} } f_{a} (\vec{p})\left(1-f_{a} (\vec{p})\right).
\end{equation}
The pole-pole contributions have the same expressions as we had in our previous investigation in the semi-classical framework [16]. The cut-cut part has four terms,
\begin{equation}\label{EqA6}
\tilde{\sigma }_{ab} (CC;\vec{k},t)=A_{ab}^{+} (\vec{k},t)+\tilde{A}_{ab}^{+} (\vec{k},t)+\tilde{A}_{ab}^{-} (\vec{k},t)+A_{ab}^{-} (\vec{k},t).
\end{equation}
The first and last terms come from the square of the first and second terms in Eq. (\ref{Eq13}), and involves double integrations over $\omega$ and $\omega'$. The diagonal and off diagonal parts in the isospin space are

\begin{equation}\label{EqA7}
{\left(\begin{array}{c} {A_{nn}^{\mp} (\vec{k},t)} \\ {A_{pp}^{\mp} (\vec{k},t)} \end{array}\right)}
 = {\int_{-\infty}^{+\infty}\frac{d\omega}{2\pi} \int_{-\infty}^{+\infty}\frac{d\omega'}{2\pi} \frac{e^{-i(\omega +\omega')t}}{\omega +\omega'\mp 2i\eta}
    \left(\begin{array}{cc} {W_{nn}^{\mp}} & {V_{np}^{\mp}} \\ {V_{pn}^{\mp}} & {W_{pp}^{\mp}} \end{array}\right) \otimes \left(\begin{array}{c} {\frac{\phi_{n} \omega\mp i\eta)+\phi_{n} (\omega'\mp i\eta)}{\varepsilon (\omega \mp i\eta )\varepsilon (\omega'\mp i\eta)}} \\ {\frac{\phi_{p}(\omega\mp i\eta)+\phi_{p} (\omega'\mp i\eta)}{\varepsilon (\omega \mp i\eta )\varepsilon (\omega'\mp i\eta)}} \end{array}\right)},
\end{equation}
 and

\begin{equation}\label{EqA8}
{A_{pn}^{\mp}(\vec{k},t)} = {-\int_{-\infty}^{+\infty}\frac{dw}{2\pi} \int_{-\infty}^{+\infty}\frac{dw'}{2\pi}
  \frac{e^{-i(\omega+\omega')t}}{\omega+\omega'\mp 2i\eta}
      \left(\begin{array}{cc} {W_{pn}^{\mp}} & {V_{nn}^{\mp}}\end{array}\right) \otimes
      \left(\begin{array}{c} {\frac{\phi_{n}(\omega\mp i\eta)+\phi_{n}(\omega'\mp i\eta)}{\varepsilon(\omega\mp i\eta)\varepsilon (\omega'\mp i\eta)}} \\ {\frac{\phi_{p} (\omega\mp i\eta)+\phi_{p} (\omega'\mp i\eta)}{\varepsilon (\omega\mp i\eta)\varepsilon (\omega'\mp i\eta)}} \end{array}\right)}.
\end{equation}
In these expressions and below, the symbol $\otimes$ denotes the matrix multiplication. The double integrals in $A_{ab}^{\mp} (\vec{k},t)$ contain the principle value and the delta function contributions, which are identified using the identity $1/\left(\omega+\omega'\mp 2i\eta \right)=P(1/\omega +\omega')\pm i\pi\delta (\omega+\omega')$. The elements of $W$ and $V$ matrices are given by

\begin{eqnarray}\label{EqA9}
\left(\begin{array}{cc}
       {W_{nn}^{\mp}} & {  }{V_{np}^{\mp}}\\
       {V_{pn}^{\mp}} & {  }{W_{pp}^{\mp}}
      \end{array} \right)
=\left(\begin{array}{cc}
      {\left(1+F_{0}^{pp}\chi_{p}^{\mp } \right)\left(1+F_{0}^{pp}{\chi'}_{p}^{\mp } \right)} &
      {{(F_{0}^{np})}^{2} \chi_{n}^{\mp } {\chi'}_{n}^{\mp } } \\
      {{(F_{0}^{pn})}^{2} \chi_{p}^{\mp } {\chi'}_{p}^{\mp } } & {\left(1+F_{0}^{nn}\chi_{n}^{\mp } \right)\left(1+F_{0}^{nn} {\chi'}_{n}^{\mp } \right)} \end{array}\right),
\end{eqnarray}
and

\begin{eqnarray}\label{EqA10}
\left(\begin{array}{c}{W_{pn}^{\mp}} \\ {V_{nn}^{\mp}}\end{array}\right)
=\left(\begin{array}{c}{F_{0}^{pn}\chi_{p}^{\mp}\left(1+F_{0}^{pp}{\chi'}_{p}^{\mp}\right)} \\
 {\left(1+F_{0}^{nn}\chi_{n}^{\mp}\right)F_{0}^{np}{\chi'}_{n}^{\mp}}\end{array}\right).
\end{eqnarray}
In these expressions and also below, we use the short hand notation $\chi_{a}^{\mp}=\chi_{a}(\vec{k},\omega\mp i\eta)$, ${\chi'}_{a}^{\mp} =\chi_{a}(\vec{k},{\omega'}\mp i\eta)$ and the quantity $\phi_{a}(\omega \mp i\eta )$ is defined as

\begin{equation}\label{EqA11}
\phi_{a}\left(\omega \mp i\eta\right)= 2\int_{-\infty}^{+\infty}\frac{d^{3}p}{(2\pi{\rm\;}\hbar)^{3}}f_{a} \left(\vec{p}\right)\left[1-f_{a}\left(\vec{p}\right)\right]\frac{1}{\vec{p}\cdot\vec{k}/m-\left(\omega \mp i\eta \right)}.
\end{equation}

There are contributions coming from the cross-terms in the square of Eq. (\ref{Eq13}), which are indicated by the second and third terms in Eq. (\ref{EqA6}). These terms also involves double integrations over $\omega$ and $\omega'$, and have similar structures as $A_{ab}^{\mp}(\vec{k},t)$

\begin{eqnarray}\label{EqA12}
     {\left(\begin{array}{c}{\tilde{A}_{nn}^{\mp}(\vec{k},t)}\\
     {\tilde{A}_{pp}^{\mp}(\vec{k},t)}\end{array}\right)} = {-\int_{-\infty}^{+\infty}\frac{d\omega}{2\pi}\int_{-\infty }^{+\infty} \frac{d{\omega'}}{2\pi}\frac{e^{-i(\omega+{\omega'})t}}{\omega+{\omega'}}
 \left(\begin{array}{cc}{\tilde{W}_{nn}^{\mp}} & {\tilde{V}_{np}^{\mp}} \\
 {\tilde{V}_{pn}^{\mp}} & {\tilde{W}_{pp}^{\mp}}\end{array}\right) \otimes
      \left(\begin{array}{c}{\frac{\phi_{n}(\omega\mp i\eta)+\phi_{n}({\omega'}\pm i\eta)}{\varepsilon(\omega\mp i\eta)\varepsilon
     ({\omega'}\pm i\eta)}} \\ {\frac{\phi_{p}(\omega\mp i\eta)+\phi_{p}({\omega'}\pm i\eta)}{\varepsilon (\omega \mp i\eta )\varepsilon ({\omega'}\pm i\eta )}} \end{array}\right)}
\end{eqnarray}
and

\begin{eqnarray}\label{EqA13}
{\tilde{A}_{pn}^{\pm}(\vec{k},t)} & {=} & {\int_{-\infty}^{+\infty}\frac{dw}{2\pi}\int _{-\infty}^{+\infty}\frac{dw'}{2\pi}  \frac{e^{-i(\omega+\omega')t}}{\omega+\omega'}
 \left(\begin{array}{cc} {\tilde{W}_{pn}^{\mp}} & {\tilde{V}_{nn}^{\mp}}\end{array}\right)  \otimes
 \left(\begin{array}{c} {\frac{\phi_{n}(\omega\pm i\eta)+\phi_{n}(\omega'\mp i\eta)}
 {\varepsilon(\omega\pm i\eta)\varepsilon (\omega'\mp i\eta )}} \\
 {\frac{\phi_{p}(\omega\pm i\eta)+\phi_{p}(\omega'\mp i\eta )}{\varepsilon(\omega\pm i\eta)\varepsilon(\omega'\mp i\eta)}} \end{array}\right)}.
\end{eqnarray}
The integrand  in $\tilde{A}_{ab}^{\mp}(\vec{k},t)$, in contrast to its appearance, is well-behaved  function, because when $\omega'$$=$$-\omega$, the nominator is also zero therefore the ratio $\left[\phi(\omega\mp i\eta)+\phi(\omega'\pm i\eta)\right]/\left(\omega +\omega'\right)$ becomes finite. Here, the elements of $\tilde{W}$ and $\tilde{V}$ matrices are given by,

\begin{eqnarray}\label{EqA14}
\left(\begin{array}{cc} {\tilde{W}_{nn}^{\mp}} & {\tilde{V}_{np}^{\mp}} \\
      {\tilde{V}_{pn}^{\mp}} & {\tilde{W}_{pp}^{\mp}}\end{array}\right)
      =\left(\begin{array}{cc} {\left(1+F_{0}^{pp}\chi_{p}^{\mp}\right)\left(1+F_{0}^{pp}{\chi'}_{p}^{\pm}\right)} &
      {(F_{0}^{np})^{2}\chi_{n}^{\mp}{\chi'}_{n}^{\pm}} \\
      {(F_{0}^{pn})^{2}\chi_{p}^{\mp}{\chi'}_{p}^{\pm}}&
      {\left(1+F_{0}^{nn}\chi_{n}^{\mp}\right)\left(1+F_{0}^{nn}{\chi'}_{n}^{\pm}\right)}\end{array}\right),
\end{eqnarray}
and

\begin{equation}\label{EqA15}
   \left(\begin{array}{c} {W_{pn}^{\mp}} \\ {V_{nn}^{\mp}}\end{array}\right)
  =\left(\begin{array}{c} {F_{0}^{pn}\chi_{p}^{\pm}\left(1+F_{0}^{pp}{\chi'}_{p}^{\mp}\right)} \\
   {\left(1+F_{0}^{nn}\chi_{p}^{\pm}\right)F_{0}^{np}{\chi'}_{n}^{\mp}} \end{array}\right).
\end{equation}

In spectral density, the pole-cut contribution has also four terms
\begin{equation}\label{EqA16}
\tilde{\sigma}_{ab}(PC;\vec{k},t)=B_{ab}^{+}(\vec{k},t)+\tilde{B}_{ab}^{+}(\vec{k},t)
+\tilde{B}_{ab}^{-}(\vec{k},t)+B_{ab}^{-}(\vec{k},t).
 \end{equation}
The isospin matrix elements of the first and the last term are given by,

\begin{equation}\label{EqA17}
   \left(\begin{array}{c} {B_{nn}^{\mp}(\vec{k},t)} \\
   {B_{pp}^{\mp}(\vec{k},t)}\end{array}\right)
   =\frac{\pm ie^{\mp \Gamma t}}{\partial \varepsilon / \partial\omega |_{\omega=\mp i\Gamma}}\int_{-\infty}^{+\infty}\frac{d\omega}
   {2\pi} \frac{e^{-i\omega t}}{\omega\mp i\Gamma}
        \left(\begin{array}{cc} {X_{nn}^{\mp}} & {Y_{np}^{\mp}} \\
         {Y_{pn}^{\mp } } & {X_{pp}^{\mp}}\end{array}\right) \otimes
   \left(\begin{array}{c} {\frac{\phi _{n}(\mp i\Gamma)+\phi_{n}(\omega\mp i\eta)}{\varepsilon(\omega\mp i\eta)}} \\
   {\frac{\phi_{p}(\mp i\Gamma)+\phi_{p}(\omega\mp i\eta)}{\varepsilon(\omega\mp i\eta)}}\end{array}\right),
\end{equation}
 and

\begin{equation}\label{EqA18}
{B_{pn}^{\mp}(\vec{k},t)} = {\frac{\mp ie^{\mp \Gamma t}}{\partial \varepsilon / \partial\omega |_{\omega=\mp i\Gamma}}
 \int_{-\infty}^{+\infty}\frac{d\omega}{2\pi} \frac{e^{-i\omega t}}{\omega\mp i\Gamma}
    \left(\begin{array}{cc} {X_{pn}^{\mp}} & {Y_{nn}^{\mp}} \end{array}\right)  \otimes
 \left(\begin{array}{c} {\frac{\phi_{n}(\mp i\Gamma_{k})+\phi_{n} (\omega\mp i\eta)}{\varepsilon(\omega\mp i\eta)}} \\
 {\frac{\phi _{p}(\mp i\Gamma_{k})+\phi_{p}(\omega\mp i\eta)}{\varepsilon(\omega\mp i\eta)}} \end{array}\right)}.
\end{equation}
Here , we use the short hand notation $\phi _{a} (\mp i\Gamma )=\phi _{a} (\vec{k},\omega =\mp i\Gamma )$, ignore the label $k$ in $\Gamma _{k}$ for simplicity, and the elements of $X$ and $Y$ matrices are given by

\begin{equation}\label{EqA19}
    \left(\begin{array}{cc} {X_{nn}^{\mp}} & {Y_{np}^{\mp}} \\
    {Y_{pn}^{\mp}} & {X_{pp}^{\mp}} \end{array}\right)=
\left(\begin{array}{cc} {\left(1+F_{0}^{pp} \chi_{p}^{\mp i\Gamma} \right)\left(1+F_{0}^{pp}\chi_{p}^{\mp}\right)} &
{(F_{0}^{np})^{2}\chi_{n}^{\mp i\Gamma}\chi_{n}^{\mp}} \\
{(F_{0}^{pn})^{2}\chi_{p}^{\mp i\Gamma}\chi_{p}^{\mp}} &
{\left(1+F_{0}^{nn}\chi_{n}^{\mp i\Gamma}\right)\left(1+F_{0}^{nn}\chi_{n}^{\mp}\right)}\end{array}\right),
\end{equation}
and

\begin{equation}\label{EqA20}
\left(\begin{array}{c} {X_{pn}^{\mp}} \\ {Y_{nn}^{\mp}} \end{array}\right)
      =\left(\begin{array}{c}{F_{0}^{pn}\chi_{p}^{\mp i\Gamma}\left(1+F_{0}^{pp}\chi_{p}^{\mp}\right)} \\
      {\left(1+F_{0}^{nn} \chi_{n}^{\mp i\Gamma}\right)F_{0}^{np}\chi_{n}^{\mp}}\end{array}\right).
\end{equation}
where $\chi _{a}^{\mp i\Gamma } =\chi _{a} (\vec{k},\omega =\mp i\Gamma )$. The second and third terms involve an integral over $\omega $, and have similar structure as $B_{ab}^{\mp} (\vec{k},t)$,

\begin{equation}\label{EqA21}
   \left(\begin{array}{c} {\tilde{B}_{nn}^{\mp} (\vec{k},t)} \\
   {\tilde{B}_{pp}^{\mp } (\vec{k},t)} \end{array}\right)
=\frac{\mp ie^{\mp\Gamma t}}{\partial\varepsilon / \partial\omega |_{\omega =\mp i\Gamma}} \int_{-\infty}^{+\infty}\frac{d\omega}{2\pi}  \frac{e^{-i\omega t}}{\omega\mp i\Gamma}
    \left(\begin{array}{cc} {\tilde{X}_{nn}^{\mp}} & {\tilde{Y}_{np}^{\mp}} \\
     {\tilde{Y}_{pn}^{\mp}} & {\tilde{X}_{pp}^{\mp}}\end{array}\right) \otimes
\left(\begin{array}{c} {\frac{\phi_{n}(\mp i\Gamma)+\phi_{n} (\omega\pm i\eta)}{\varepsilon(\omega\pm i\eta)}} \\
{\frac{\phi _{p} (\mp i\Gamma)+\phi_{p}(\omega\pm i\eta)}{\varepsilon (\omega \pm i\eta)}} \end{array}\right),
\end{equation}
and

\begin{equation}\label{EqA22}
{\tilde{B}_{pn}^{\mp} (\vec{k},t)} = {\frac{\pm ie^{\mp\Gamma t}}{\partial\varepsilon /\partial\omega |_{\omega=\mp i\Gamma}}
 \int_{-\infty}^{+\infty}\frac{d\omega}{2\pi} \frac{e^{-i\omega t}}{\omega\mp i\Gamma}
  \left(\begin{array}{cc} {\tilde{X}_{pn}^{\mp}} & {\tilde{Y}_{nn}^{\mp}} \end{array}\right) \otimes
   \left(\begin{array}{c} {\frac{\phi_{n} (\mp i\Gamma)+\phi _{n} (\omega\pm i\eta)}{\varepsilon (\omega\pm i\eta)}} \\
   {\frac{\phi_{p}(\mp i\Gamma)+\phi_{p}(\omega\pm i\eta)}{\varepsilon (\omega\pm i\eta)}} \end{array}\right)}.
\end{equation}
The elements of $\tilde{X}$ and $\tilde{Y}$ matrices are given by

\begin{equation}\label{EqA23}
\left(\begin{array}{cc} {\tilde{X}_{nn}^{\mp } } & {\tilde{Y}_{np}^{\mp } } \\ {\tilde{Y}_{pn}^{\mp } } & {\tilde{X}_{pp}^{\mp } } \end{array}\right)=\left(\begin{array}{cc} {\left(1+F_{0}^{pp} \chi _{p}^{\mp i\Gamma } \right)\left(1+F_{0}^{pp} \chi _{p}^{\pm } \right)} & {(F_{0}^{np} )^{2} \chi _{n}^{\mp i\Gamma } \chi _{n}^{\pm } } \\ {(F_{0}^{pn} )^{2} \chi _{p}^{\mp i\Gamma } \chi _{p}^{\pm } } & {\left(1+F_{0}^{nn} \chi _{n}^{\mp i\Gamma } \right)\left(1+F_{0}^{nn} \chi _{n}^{\pm } \right)} \end{array}\right),
\end{equation}
and

\begin{equation}\label{EqA24}
\left(\begin{array}{c} {\tilde{X}_{pn}^{\mp } } \\ {\tilde{Y}_{nn}^{\mp } } \end{array}\right)=\left(\begin{array}{c} {F_{0}^{pn} \chi _{p}^{\mp i\Gamma } \left(1+F_{0}^{pp} \chi _{p}^{\pm } \right)} \\ {\left(1+F_{0}^{nn} \chi _{n}^{\mp i\Gamma } \right)F_{0}^{np} \chi _{n}^{\pm } } \end{array}\right).
\end{equation}

\begin{figure}[ht]
\hspace{1.0cm}
\includegraphics[width=11cm, height=21cm]{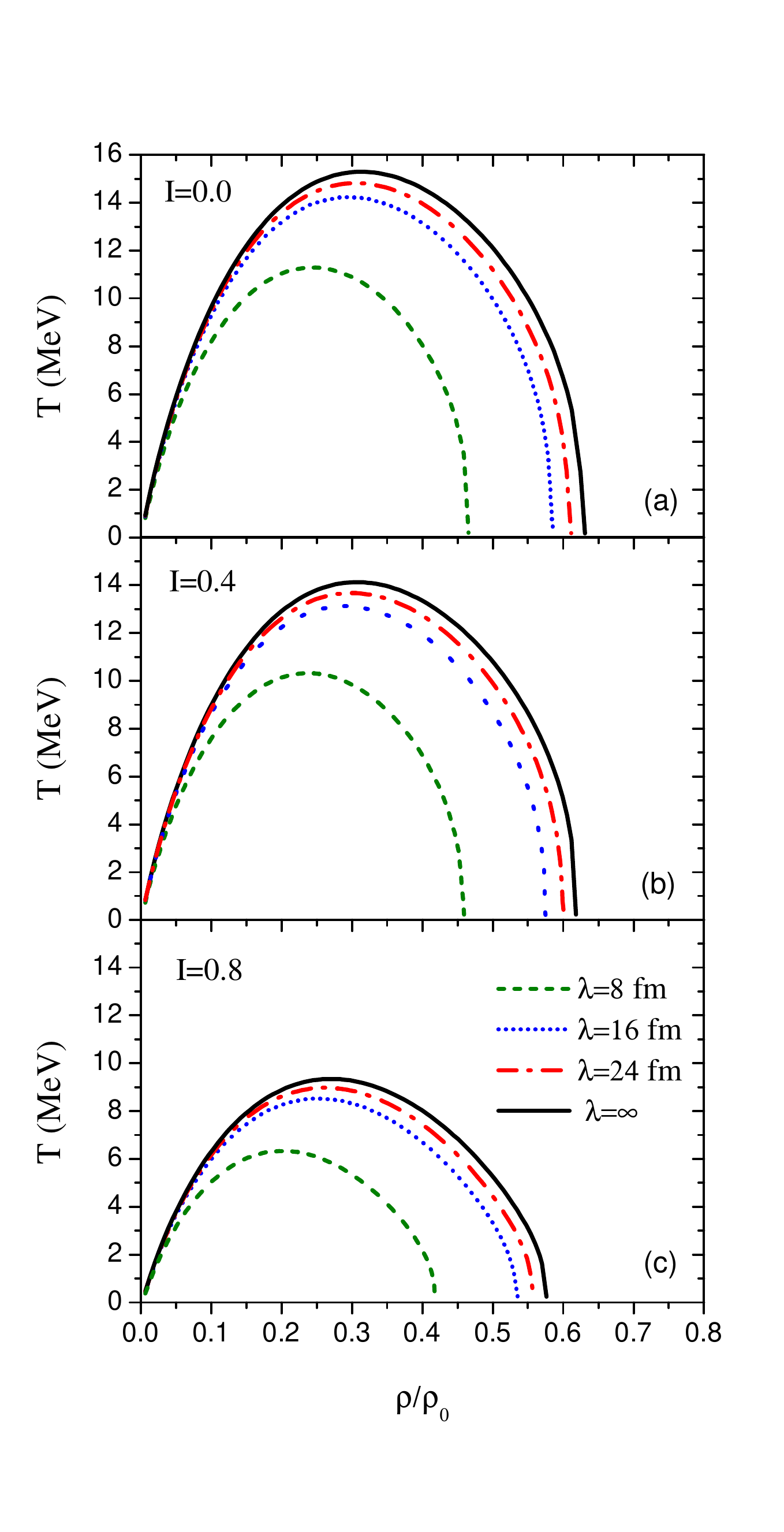}
\vspace{0.0cm}
 \caption{\label{fig1}(Color online) Phase diagram in density-temperature plane for different wavelengths corresponding to the potential given by Eq. (\ref{Eq19}).}
\vspace{0.0cm}
\end{figure}

\begin{figure}[b]
\hspace{-0.5cm}
\includegraphics[width=11cm, height=21cm]{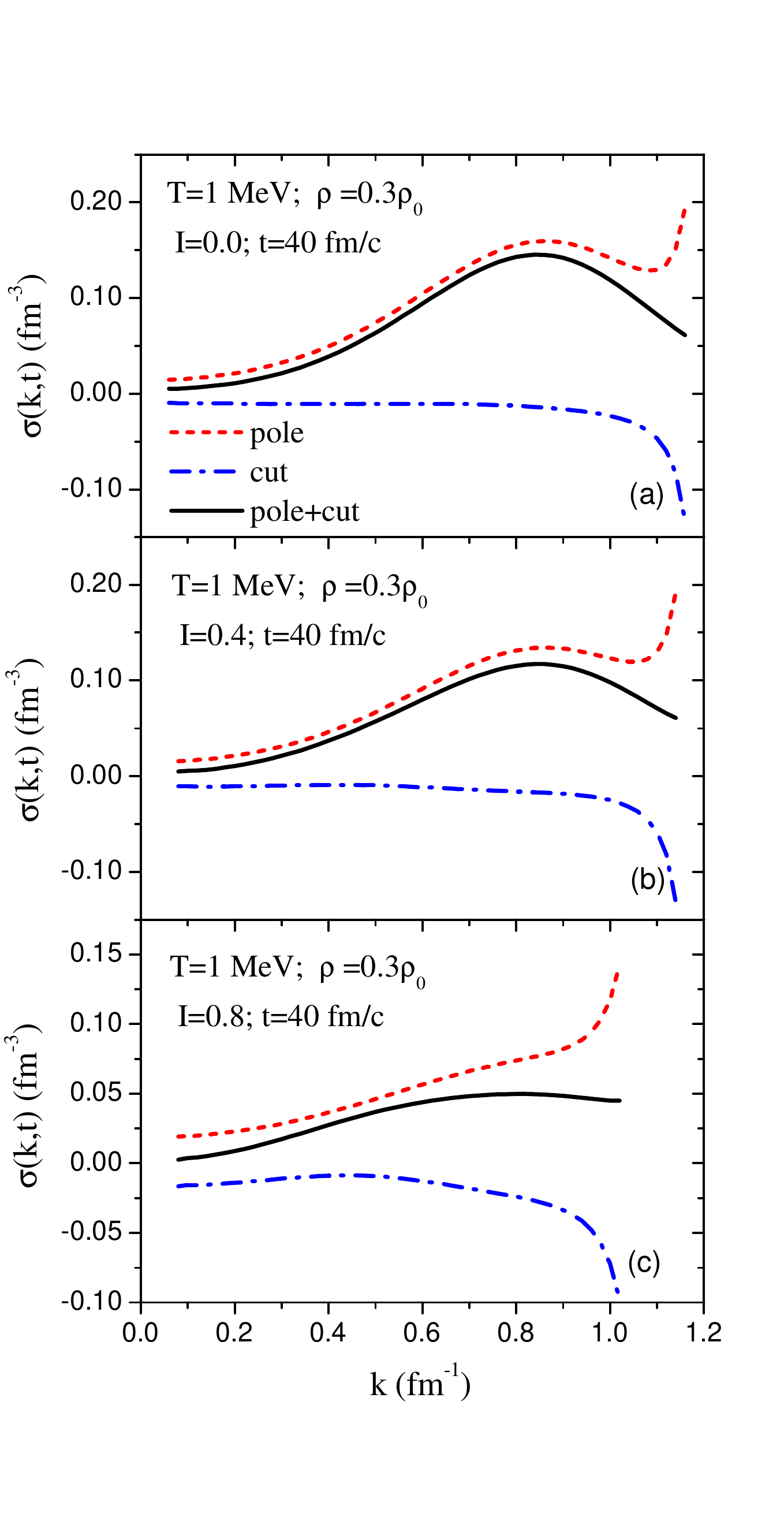}
\vspace{-0.5cm}
 \caption{\label{fig2}(Color online) Spectral intensity of the correlation function as a function of wave number at initial density $\rho =0.3\rho_{0}$ ${\text{fm}}^{-3}$ at time $t=40$ fm/$c$ at temperature $T=1$ MeV for three different charge asymmetries. Dotted, dashed-dotted and solid lines are results of pole, cut and total contributions, respectively.}
\vspace{-0.0cm}
\end{figure}

\begin{figure}[h]
\hspace{-0.5cm}
\includegraphics[width=11cm, height=21cm]{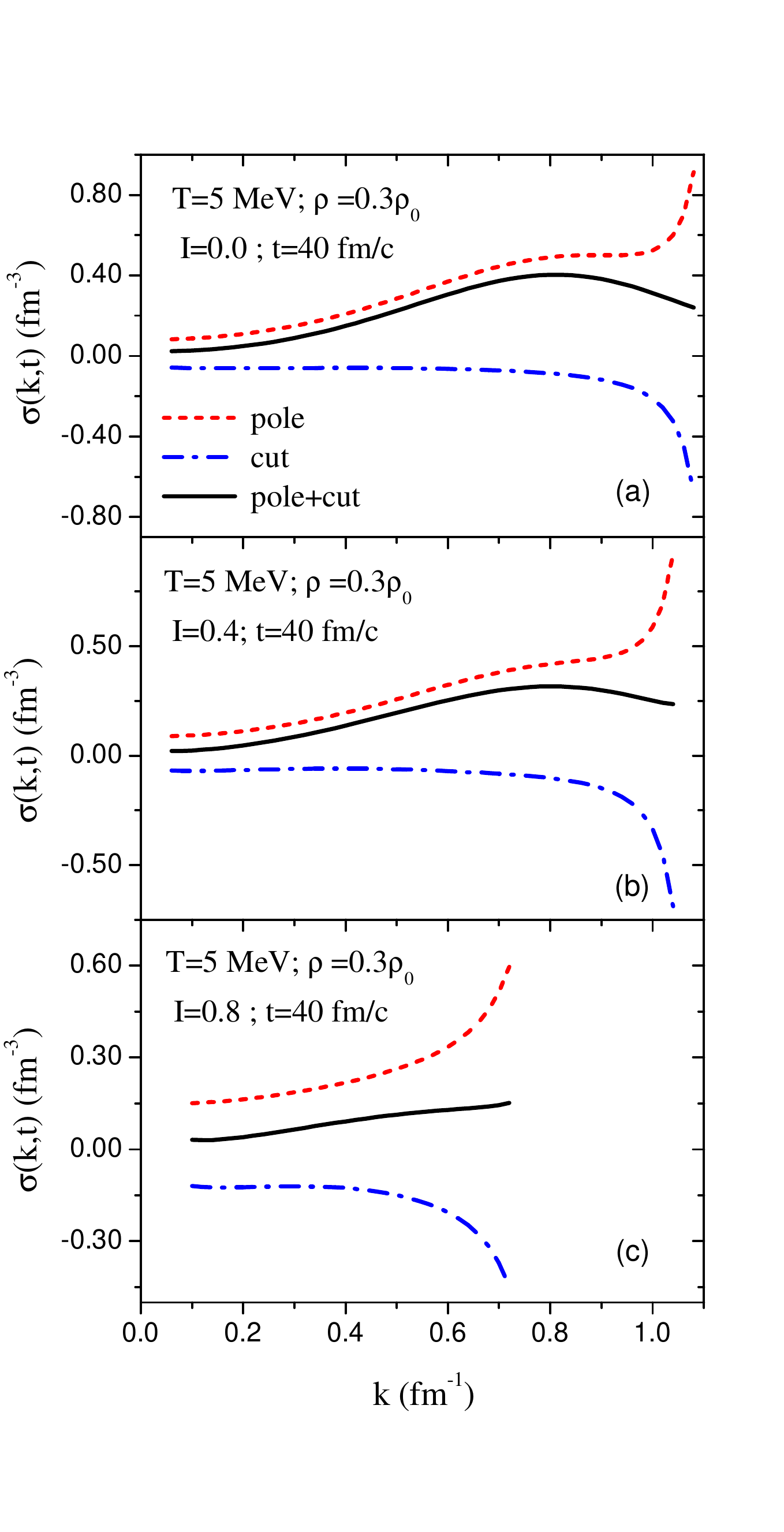}
\vspace{-1.5cm}
 \caption{\label{fig3}(Color online) Spectral intensity of the correlation function as a function of wave number at initial density $\rho =0.3\rho _{0}$ ${\text{fm}}^{-3} $ at time $t=40$ fm/$c$ at temperature $T=5$ MeV for three different charge asymmetries. Dotted, dashed-dotted and solid lines are results of pole, cut and total contributions, respectively. }
\vspace{-0.0cm}
\end{figure}

\begin{figure}[h]
\hspace{-0.5cm}
\includegraphics[width=11cm, height=14cm]{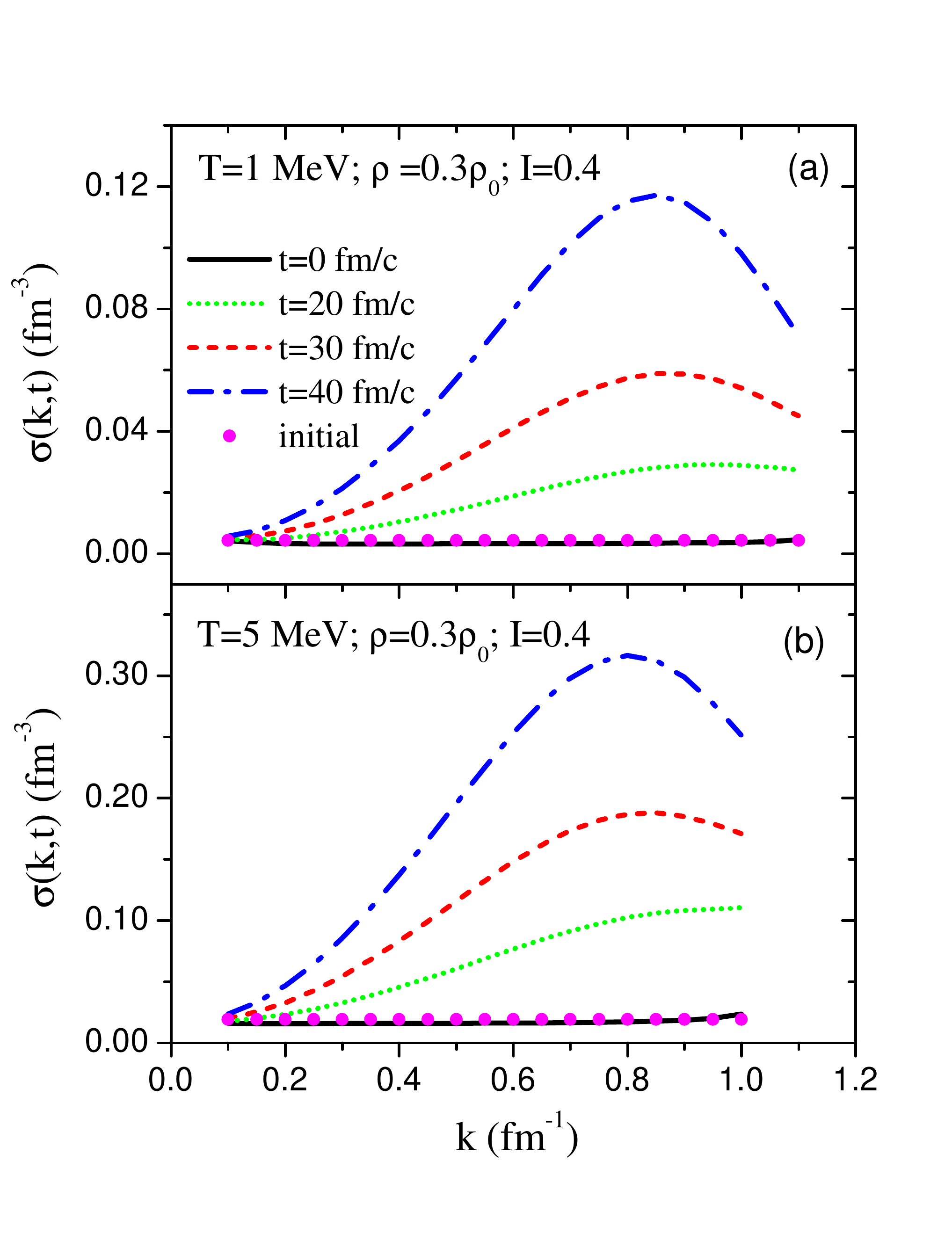}
\vspace{-1.5cm}
 \caption{\label{fig4}(Color online) Spectral intensity of the correlation function as a function of wave number at initial density $\rho =0.3\rho _{0}$ ${\text{fm}}^{-3} $ and charge asymmetry $I=0.4$ for different times at temperature $T=1$ MeV (a) and $T=5$ MeV (b). Dots on the solid lines at times $t=0$ represent the initial conditions.}
\vspace{-0.0cm}
\end{figure}

\begin{figure}[h]
\hspace{-0.5cm}
\includegraphics[width=11cm, height=21cm]{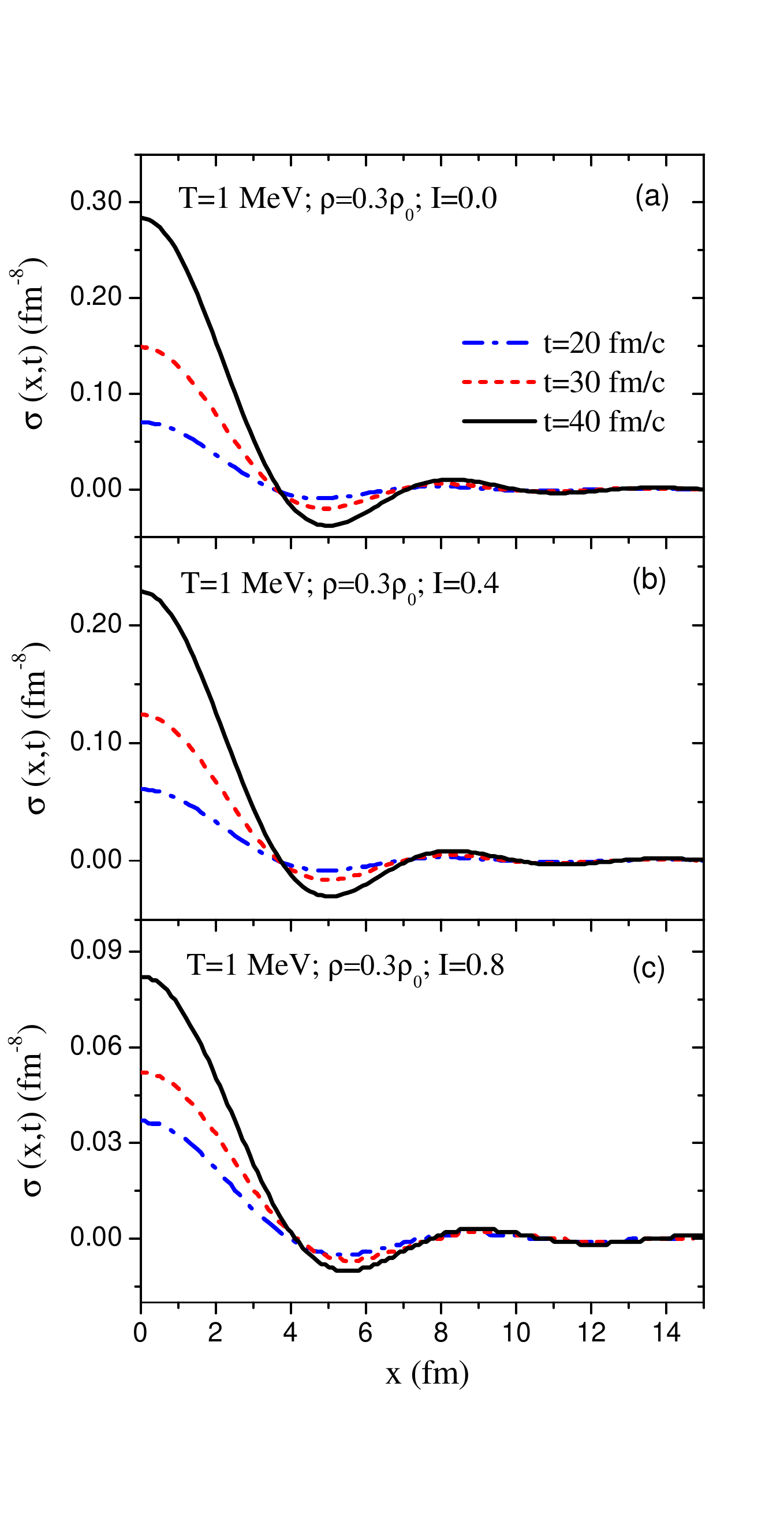}
\vspace{-1.5cm}
 \caption{\label{fig5}(Color online)Density correlation function as a function of distance between two space location $x=|\vec{r}-\vec{r}'|$ for initial density $\rho =0.3\rho_{0}$ ${\text{fm}}^{-3}$ and different charge asymmetries at temperature $T=1$ MeV at times $t=20,30,40$ fm/$c$. }
\vspace{-0.0cm}
\end{figure}

\begin{figure}[h]
\hspace{-0.5cm}
\includegraphics[width=11cm, height=21cm]{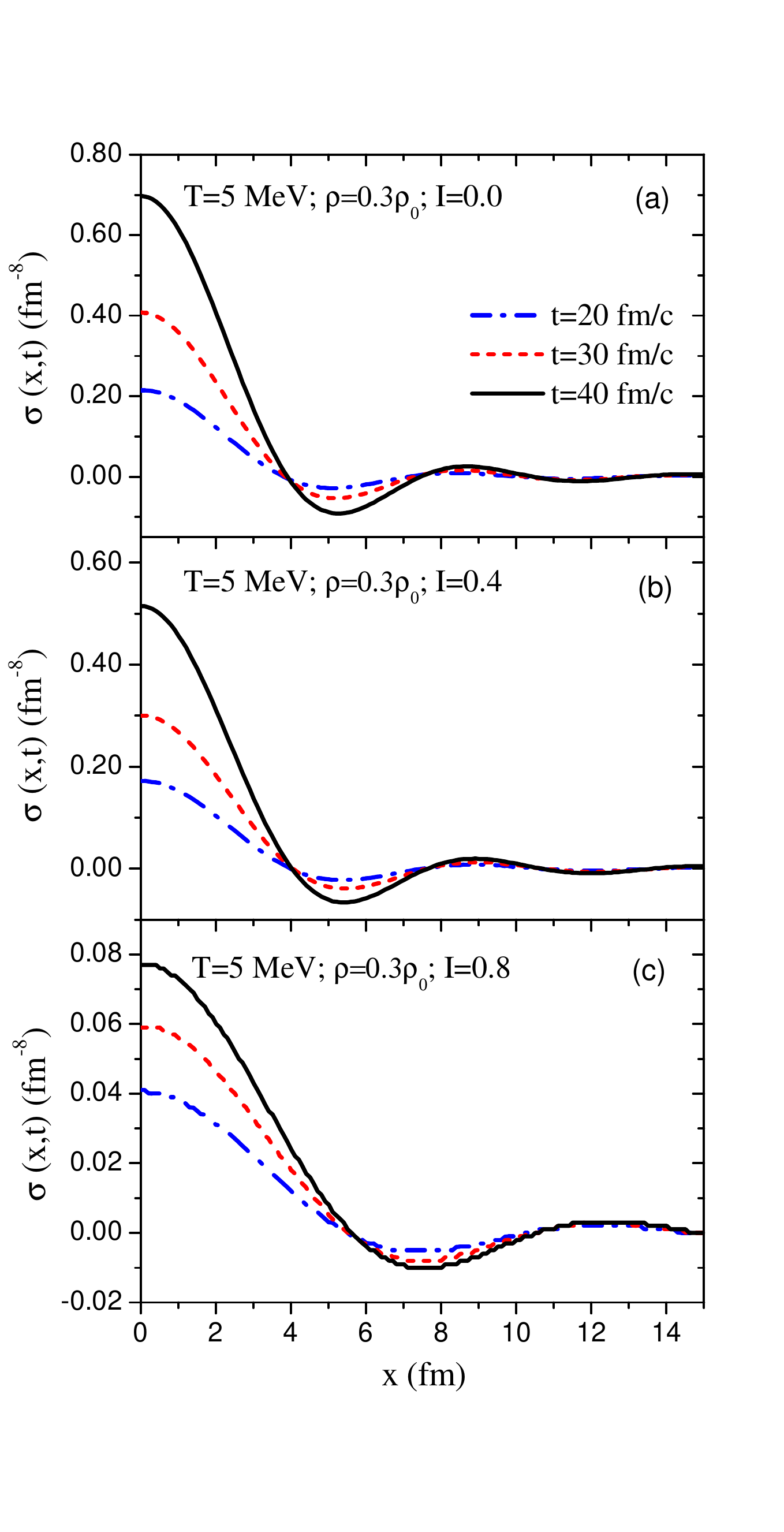}
\vspace{-1.5cm}
 \caption{\label{fig6}(Color online)  Density correlation function as a function of distance between two space location $x=|\vec{r}-\vec{r}'|$ for initial density $\rho =0.3\rho_{0}$ ${\text{fm}}^{-3}$ and different charge asymmetries at temperature $T=5$ MeV at times $t=20,30,40$ fm/$c$.}
\vspace{-0.0cm}
\end{figure}

\end{document}